\documentclass[lettersize,journal]{IEEEtran}
\usepackage{amsmath,amsfonts}
\usepackage{algorithmicx}
\usepackage{array}
\usepackage[caption=false,font=normalsize,labelfont=sf,textfont=sf]{subfig}
\usepackage{textcomp}
\usepackage{stfloats}
\usepackage{url}
\usepackage{verbatim}
\usepackage{graphicx}
\usepackage{float}
\usepackage{multirow}
\usepackage{cite}
\usepackage{amsmath,amssymb,amsfonts}

\usepackage{url}
\usepackage{hyperref}
\usepackage{graphicx}
\usepackage{textcomp}
\usepackage{adjustbox}
\usepackage{diagbox}
\usepackage{xcolor}
\usepackage[utf8]{inputenc}
\usepackage{slashbox}
\usepackage{fourier} 
\usepackage{array}
\usepackage{makecell}
\usepackage{comment}
\usepackage[para,online,flushleft]{threeparttable}
\usepackage{algpseudocode}
\usepackage{booktabs}
\usepackage{adjustbox,lipsum}
\usepackage{placeins}
\makeatletter
\newcommand{\linebreakand}{%
  \end{@IEEEauthorhalign}
  \hfill\mbox{}\par
  \mbox{}\hfill\begin{@IEEEauthorhalign}
}
\makeatother

\usepackage[
  separate-uncertainty = true,
  multi-part-units = repeat
]{siunitx}

\def\BibTeX{{\rm B\kern-.05em{\sc i\kern-.025em b}\kern-.08em
    T\kern-.1667em\lower.7ex\hbox{E}\kern-.125emX}}

\usepackage{balance}
\usepackage{titlesec}

\titleformat{\paragraph}
{\normalfont\normalsize\bfseries}{\theparagraph}{1em}{}
\titlespacing*{\paragraph}
{0pt}{3.25ex plus 1ex minus .2ex}{1.5ex plus .2ex}

\begin{document}
\title{Auto-calibrated Wearable System for Load Vertical Location Estimation During Manual Lifting}
\author{{ Diliang Chen, Nozhan Ghoreishi, John LaCourse, Sajay Arthanat, Dain LaRoche}

\thanks{This research was funded by the University of New Hampshire, Collaborative Research Excellence (CoRE) Initiative, Pilot Research Partnerships (PRPs), and the UNHInnovation fund.}}

\maketitle
\begin{abstract}
Lifting during manual material handling is a major cause of low-back pain (LBP). As an important risk factor that directly influences the risk of LBP, the Load vertical location (LVL) during lifting needs to be measured and controlled. However, existing solutions for LVL measurement are inefficient, inaccurate, and impractical for real-world workplace environments. To address these problems, an unobtrusive wearable system, including smart insoles and smart wristbands, was proposed to measure LVL accurately in workplace environments. 
Different from traditional methods which rely on Inertial Measurement Unit (IMU) and suffer from integral drifting errors over time, a novel barometer-based LVL measurement method was proposed in this study. To correct the environment-induced LVL measurement errors in the barometer-based method, a novel Known Vertical Location Update (KVLU) method was proposed. This method calibrates the measured LVL using a known wrist vertical location at known postures during frequently used daily activities such as standing and walking. The proposed wearable system achieved a mean absolute error (MAE) of 5.71 cm in LVL measurement. This result indicates that the proposed system has the potential to reliably measure LVL and assess the risk of LBP in manual lifting tasks. 


 \end{abstract}


\section{Introduction}
Low-back pain (LBP) is one of the most prevalent health problems in workplaces that adds a significant burden on the US healthcare system\cite{nih1}. 
In 2016, LBP resulted in a direct expenditure of about \$135 billion for the US healthcare system \cite{healthdata}. LBP also accounted for 38.5\% of work-related health problems in the same period\cite{BLS4}. 
Manual lifting is a major cause of LBP\cite{kingma}. According to the Bureau of Labor Statistics, manual lifting contributes to 75\% of LBP \cite{occupational80back}. Therefore, identifying lifting hazards in manual lifting will be important to the control of LBP.

Load vertical location (LVL) is one of the most important parameters for evaluating the risk of LBP. Recent studies indicate that the tension of the multifidus muscle, a key muscle in the lumbar spine, is significantly influenced by the LVL during lifting activities. Additionally, there is a correlation between elevated tension in the back muscles during physical tasks such as lifting, and an increased risk of LBP development \cite{fang2021effects, helfenstein2010occupational}. In addition, LVL is used by the existing standard tool recommended by the National Institute of Safety and Health (NIOSH) -- Revised NIOSH Lifting Equation (RNLE), for assessing the injury risks during manual lifting \cite{waters1994applications}. Therefore, accurate measurement of LVL in manual lifting is critical for assessing exposure to lifting hazards and mitigating the risk of LBP.

However, existing methods for measuring LVL are limited in efficiency, accuracy, and user-friendliness, especially in uncontrolled workplace environments. Observational methods conducted by professional ergonomists are popularly used in practice for lifting risk assessment \cite{6}. However, these methods are time-consuming and prone to subjective errors \cite{7}. Computer-vision-based approaches can track LVL accurately \cite{8}. However, these methods are susceptible to complex lighting conditions and obstacles, which make them impractical in uncontrolled workplace environments \cite{yunus, valero2016musculoskeletal}. Wearable technologies are promising in measuring LVL in uncontrolled workplaces because they enable continuous measurement and are less affected by environmental factors like light and obstacles. However, the popularly used Inertial Measurement Unit (IMU) based wearable solutions suffer from integral drifting errors over time \cite{barim, hlucny2020characterizing}. In addition, IMU-based solutions usually require the user to wear numerous sensors on different body segments which hinders user adoption in real workplace settings.
In contrast, the barometer sensor can avoid the shortages of IMU because it does not have integral drifting errors in LVL measurement and requires only up to two barometer sensors for LVL tracking \cite{ghoreishi2023barometer}. The barometer sensor has been applied in various applications involving vertical displacement. 
For example, barometer sensors embedded inside mobile phones were used to classify between standing/walking and riding an elevator \cite{vanini2016using}, and differentiate dynamic activities such as ascending or descending stairs and static activities such as sitting \cite{xie2018human}. In our previous study, the barometer sensor was used for estimating wrist vertical travel distance by measuring the air pressure changes during each short-term lifting \cite{ghoreishi2023barometer}. 
However, the barometer's measurements cannot be directly used to provide a reliable estimate of LVL over a long time (e.g. 8 hours) because of the environment-induced drifting errors \cite{matyja2022mems}. 

To address the problems in LVL measurement, an auto-calibrated and unobtrusive wearable system, consisting of a pair of smart insoles and a pair of smart wristbands, was proposed to measure LVL accurately in workplace environments. To correct the barometer drifting problem over time, a novel auto-calibration method was proposed by synergistically fusing human gait, posture, and anthropometry data. Inspired by the well-known Zero Velocity Update (ZUPT) method which significantly reduces the drift errors of the shoe-attached accelerometer in measuring gait velocity by updating the velocity with the known zero velocity during the foot flat phase, a Known Vertical Location Update (KVLU) calibration method was proposed. This calibration method can effectively remove drift errors in LVL measurement by updating the estimated wrist vertical location with a known value at recognized calibration postures by the wearable system. 

The contributions of this research are:
\begin{itemize}
\item Developed an unobtrusive wearable system that can measure LVL reliably over the long term in uncontrolled workplace environments. Different from existing wearable system designs that require sensor attachment to various body segments (e.g the back) unaccustomed to additional accessories, the proposed wearable system only integrated sensing devices into everyday used insole and well-accepted wristbands to enhance the user acceptance of the system.
\item Proposed an auto-calibration method -- KVLU, by synergistically fusing human gait, posture, and anthropometry data to effectively correct drifting errors during LVL measurement. The KVLU auto-calibration method enables the wearable system to accurately measure LVL over long periods.
\item Evaluated the performance of the proposed wearable system through a comprehensive experiment involving lifting activities on various vertical levels.

\end{itemize}

\section{Related Works}

Various methods have been proposed for automatic LVL measurement. According to David et al. \cite{david2005ergonomic}, the current methods for measuring LVL can be categorized into two groups: 1) computer-vision-based methods and 2) wearable-based methods. 

For computer-vision-based LVL measurement methods, cameras are either used individually or combined with markers. Marker-based methods involve installing optical markers on workers' body landmarks. The 3D coordinates of the markers can be measured using the cameras installed in the space. 
Zhao et al. \cite{zhao2022ergonomics} used 25 markers attached to various parts of participants' bodies and 8 Osprey cameras to track participants' movements and evaluate the ergonomic risk of load raising and lowering. LVL can be estimated based on the calculated vertical location of the wrist during lifting. 
Marker-based systems are recognized as a gold standard for analyzing movements during lifting \cite{patrizi2016comparison}.  
To simplify measurement implementation, marker-less technologies have been proposed. 
Marker-less methods rely on cameras only to record the workers' movement during lifting. Computer vision algorithms are used to build the 3D skeletal model of the worker and extract 3D coordinates of workers' body segments from the recorded videos. 
Mehrizi et al. \cite{mehrizi2018computer} used two digital cameras to record the subjects doing lifting tasks from views of 90 degrees (side view) and 135 degrees and achieved an accuracy comparable to marker-based motion tracking systems. 
Although computer-vision-based methods can achieve high accuracy in LVL measurement, they are susceptible to complex lighting conditions and obstacles, which make them impractical in uncontrolled workplace environments \cite{yunus, valero2016musculoskeletal}. In addition, computer-vision-based methods involve using expensive camera systems which are not cost-efficient for LVL measurement across a large group of workers.

Wearable systems can avoid the shortages of computer-vision-based methods because they are low-cost and less susceptible to environmental factors like light and obstacles. 
Most wearable-system-based approach uses direct measurements from wearable sensors such as the IMUs for evaluating LVL. 
For example, Hlucny et al. \cite{hlucny2020characterizing} used the Xsens Link IMU motion capture system to record participants' movement while doing symmetric and asymmetric listings tasks. Seventeen IMU sensors were attached to various segments of the participants' bodies, and Xsens MVN software was used to build the body biomechanical model and extract segment orientation and location from the IMU data. LVL was estimated as the vertical height of the mid-point between two wrists during lifting, and on average, within 25 cm of the reference values. 
However, attaching numerous sensors to the workers' bodies can restrict their movements and cause pragmatic barriers to the system's adoption in workplaces \cite{zhang2022manufacturing}. 
To address this limitations, Barim et al. \cite{barim} improved the wearable system by using 5 IMU sensors attached to wrists, thighs, and waist to measure LVL. Segment angles were extracted from the IMU attached to the respective segments. During lifting, these segment angles were combined with segment lengths to estimate LVL. However, LVL estimated using this method has a large mean error of 33 cm. 
The low accuracy of the existing wearable-based method is potentially due to the integration error in angular measurement, magnetic interference, and the accumulated error across related body segments needed for LVL measurement.

To address the shortcomings of existing solutions, a user-friendly wearable system was proposed in this paper for LVL measurement. This system used barometers instead of IMUs for LVL measurement to avoid integration and accumulation errors. Additionally, a novel auto-calibration method was proposed to ensure LVL measurement accuracy over the long-term. 

\section{Methods}

In this section, details of the unobtrusive wearable system design and the KVLU method for LVL calibration were specified.

\subsection{Wearable System Design}
\begin{figure}[h]
\begin{center}
  \includegraphics[width=0.3\textwidth]{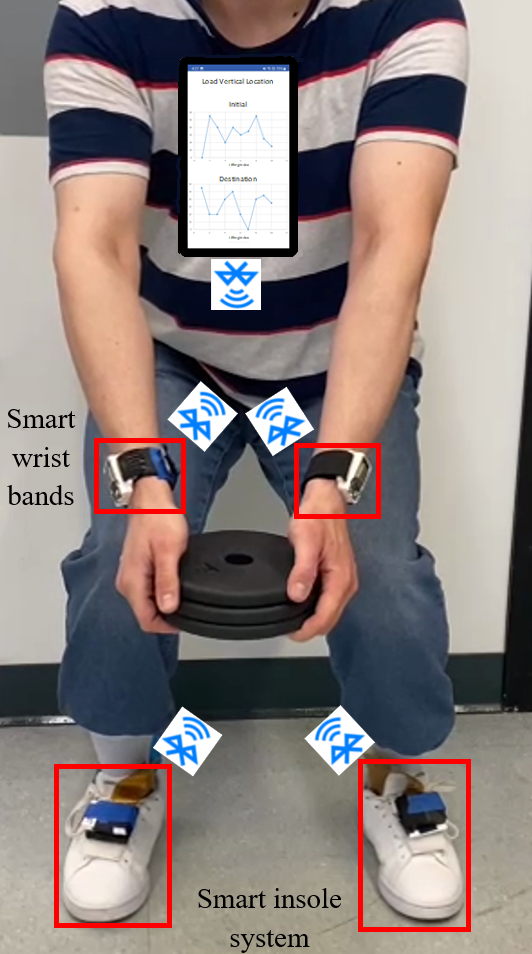}
  \caption{An overview of the unobtrusive wearable sensing system for reliable LVL measurement.}
  \label{WearableSystemOverview}
  \end{center}
\end{figure}

Fig. \ref{WearableSystemOverview} shows an overview of the proposed wearable system for accurate LVL measurement. The system consists of three parts: a pair of smart wristbands, a pair of smart insole systems, and a smartphone application. The smart wristband measures air pressure and wrist motion, which are critical for LVL measurement and calibration. In this study, the LVL is measured by tracking the wrist’s vertical location because the load lifted with the hands is close to the wrists. The smart insole system measures plantar pressure and foot motion, which are critical for recognizing activities for calibration and work-related activities such as standing, walking, and lifting \cite{chen2017risk, chen2018risk, chen2019bring}. The smartphone application is used for real-time collection and analysis of all the sensor data transmitted from the smart wristband and the smart insole via Bluetooth. 

\begin{figure}[h]
\begin{center}
  \includegraphics[width=0.4\textwidth]{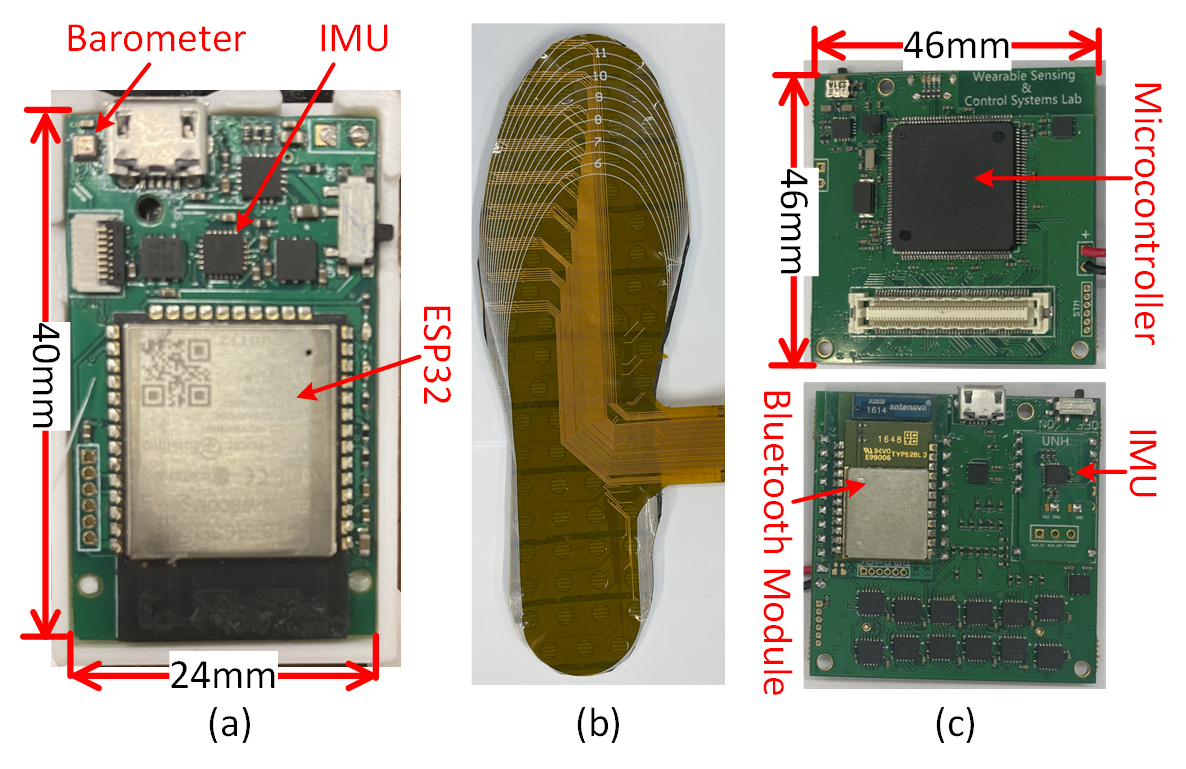}
  \caption{Hardware details of the unobtrusive wearable system. (a) Hardware dimension and electronic components of the smart wristband; (b) the flexible pressure sensor array of the smart insole; and (c) the PCB dimension and electronic components of the smart insole.}
  \label{WearableSystemDetails}
  \end{center}
\end{figure}

Hardware details of the smart wristband and smart insole systems are shown in Fig. \ref{WearableSystemDetails}. As shown in Fig. \ref{WearableSystemDetails} (a), the smart wristband consists of a barometer for measuring air pressure, an IMU for measuring wrist motion, and an ESP32 for collecting the sensor data from the barometer and IMU and transmitting the data to the smartphone application via Bluetooth. The LVL was measured by tracking the wrist's vertical location with the smart wristband during the lifting. The dimensions of the smart wristband are 40 mm x 24 mm, which makes it easy to wear on the wrist. 

The smart insole consists of a flexible pressure sensor array for measuring plantar pressure (Fig. \ref{WearableSystemDetails} (b)) and a Printed Circuit Board (PCB, Fig. \ref{WearableSystemDetails} (c)) integrated with an IMU for foot motion measurement, a microcontroller for collecting the sensor data from the flexible pressure sensor array and the IMU, and a Bluetooth module for transmitting sensor data to the smartphone application for further processing. The pressure sensor array has up to 96 pressure sensors uniformly distributed on it. To optimize the fitness of the flexible pressure sensor array for various foot sizes, an one-size-fit-all design is applied to make the pressure sensor array trimmable to fit different sizes of foot \cite{chen2018customizable}. The current insole design can be trimmed to fit foot sizes from 5.5 US to 14 US. The dimensions of the PCB are 46 mm x 46 mm, which makes it easy to be fixed on the shoelaces. 

\subsection{LVL calibration with KVLU}
Our previous research has shown that the barometer in the smart wristband can measure the changes of LVL during lifting by tracking the air pressure changes. However, due to the environmental introduced drift errors, the air pressure measured by the barometer cannot be directly used to estimate the LVL accurately. To address this problem, the KVLU method was proposed to calibrate the LVL drift errors with known wrist vertical locations at KVLU reference points.

\begin{figure}[h!]
\begin{center}
  \includegraphics[width=0.4\textwidth]{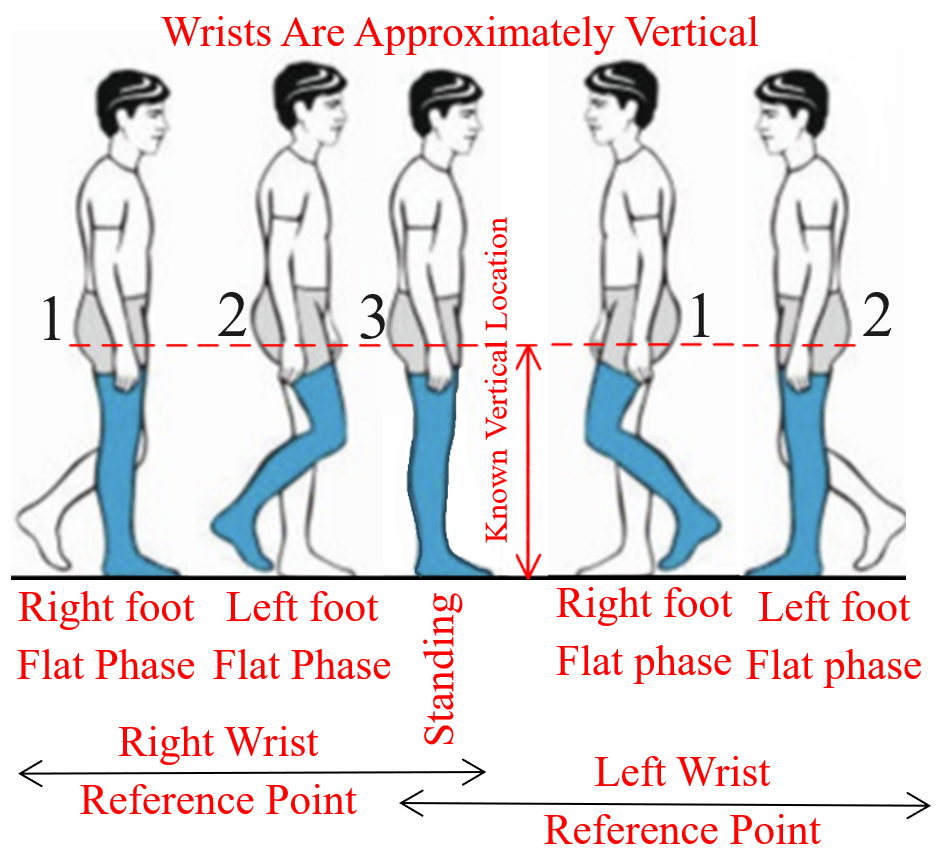}
  \caption{The KVLU reference points with a known vertical location for calibrating the LVL measurement.}
  \label{VerticalReferenceLocation}
  \end{center}
\end{figure}


Fig. \ref{VerticalReferenceLocation} shows the known vertical location used for LVL calibration. The KVLU reference points with known vertical location can be detected when the foot is flat, leg is straight, and the wrist is positioned vertically. Since the length of different body segments can be estimated based on the known body height, the known vertical location (height) of the KVLU reference points can be estimated with the known body height \cite{dongre2021correlation}. 
The KVLU reference points were identified in three postures: (1) standing with wrist(s) in the vertical posture; (2) the foot flat phase of the right foot during walking with wrist(s) in a vertical posture; and (3) the foot flat phase of the left foot during walking with wrist(s) in a vertical posture.
Since standing and walking are among the most common activities performed by workers, the identified KVLU reference points can be used to frequently calibrate LVL measurement errors.

\subsubsection{KVLU Reference Point Identification During Standing}
\begin{figure}[h!]
\begin{center}
  \includegraphics[width=0.3\textwidth]{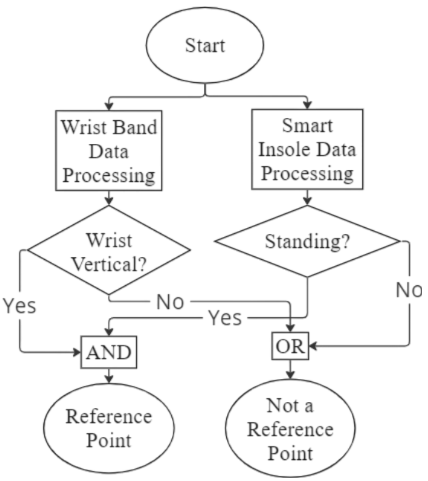}
  \caption{Flowchart of the proposed method for detecting KVLU reference points during standing}
  \label{StandingRefFlowChart}
  \end{center}
\end{figure}

The process to identify KVLU reference points during standing is shown in Fig. \ref{StandingRefFlowChart}. The KVLU reference points are identified when the wrist is in a vertical posture and the user is standing. Standing posture can be accurately recognized with the smart insole system \cite{chen2019bring}. The IMU in the smart wristband was used to determine whether the wrist was in a vertical posture. Determined by the IMU placement in the smart wristband, the measured pitch angle of the wrist increases as the wrist gets closer to the vertical position and decreases as the wrist posture deviates from it. However, during normal standing with wrists resting at the sides, the wrists are not perfectly vertical. Typically, the wrist pitch angle exhibits small variability because of subtle body movement and inter-subject differences in standing postures. Therefore, an angular threshold was used to determine if the wrists were in a suitable ``vertical'' posture.

We hypothesize that the distribution of the wrist pitch angle at the KVLU reference points during normal standing fits a Gaussian distribution. According to the empirical rule of the Gaussian distribution, nearly all the data (99.73\%) would lie within three standard deviations of the mean.
Therefore, to determine whether the wrist is in a suitable vertical posture, a threshold for the wrist pitch angle was calculated with Eq. \Ref{eq1}. 
\begin{equation}
Threshold_{angle} = \frac{1}{n_{angle}}{\sum_{k=1}^{n_{angle}} angle - 3 * \sigma_{angle}}
\label{eq1}
\end{equation}

Where, $n_{angle}$ indicates the number of wrist angle samples collected at the KVLU reference point during standing, and $\sigma_{angle}$ is the standard deviation of those wrist angle samples.
Only when the measured wrist angle is larger than the $Threshold_{angle}$, the wrist is recognized as in a suitable ``vertical'' posture. 



\subsubsection{KVLU Reference Point Identification During Walking}

\begin{figure}[h!]
\begin{center}
  \includegraphics[width=0.3\textwidth]{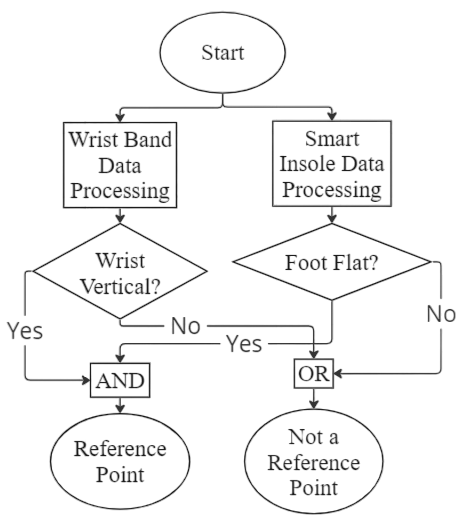}
  \caption{Flowchart of the proposed method for detecting KVLU reference point during walking}
  \label{WalkingRefPointDetection2}
  \end{center}
\end{figure}

The process for identifying KVLU reference points during walking is shown in Fig. \ref{WalkingRefPointDetection2}. The KVLU reference points are identified when the wrist is in a vertical posture and the foot is in the foot flat phase of the walking activity. Walking can be accurately recognized using the smart insole system \cite{chen2019bring}. Since both the heel and forefoot will be in contact with the ground during the foot flat phase, the flexible pressure sensor array in the smart insole is used to recognize the foot flat phase by detecting if the heel and forefoot areas are actively pressed. To accurately determine if the heel or the forefoot area was pressed, dynamic thresholds were calculated using the following equation Eq. \ref{eq_Heel} and \ref{eq_forefoot}. 

\begin{equation} 
Threshold_{heel} = 
\frac{1}{n_{swing}}{\sum_{k=1}^{n_{swing}} HeelGRF_k + 3 * \sigma_{heel}} 
\label{eq_Heel}
\end{equation}

\begin{equation} 
Threshold_{fore} = 
\frac{1}{n_{swing}}{\sum_{k=1}^{n_{swing}} ForeGRF_k + 3 * \sigma_{fore}} 
\label{eq_forefoot}
\end{equation}

\begin{figure*}[ht!]
\begin{center}
  \includegraphics[width=\textwidth]{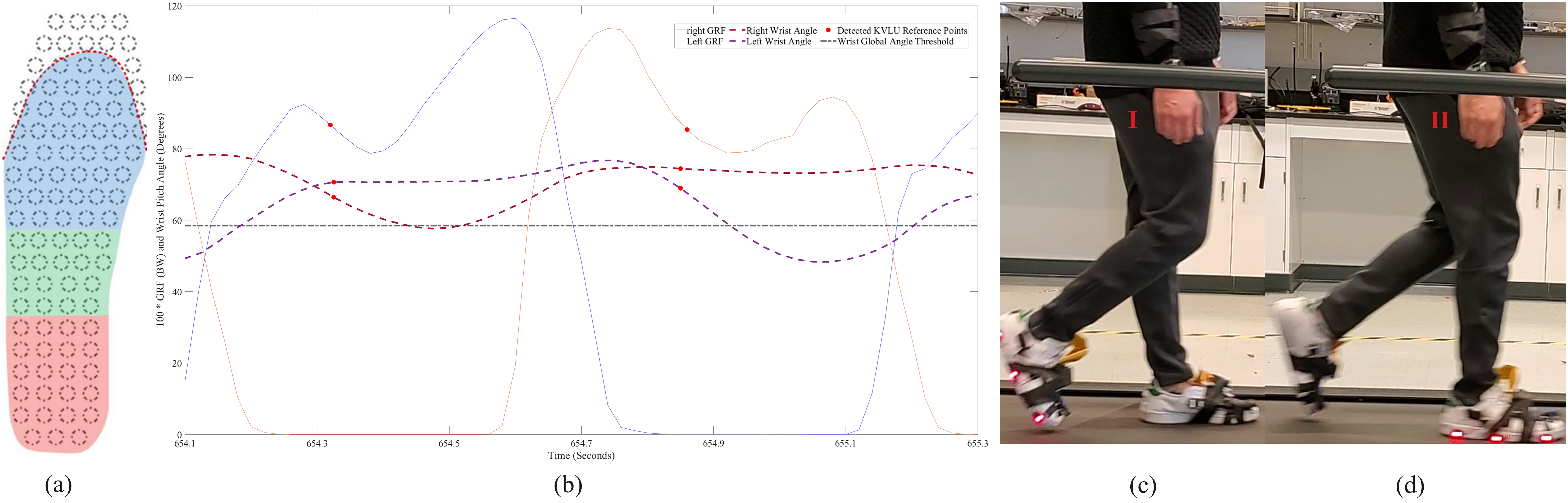}
  \caption{Details of the calibration reference point based on walking activity: (a). Different foot regions used for detecting the foot flat phase of the gait cycle. Each Red, green, and red mark the heel, mid-foot, and fore-foot area respectively. The red dashed line delimits the 10.5 US insole size. Each hollow circle represents one pressure sensor of the pressure sensor array. (b). The total GRF, as well as wrist angles during one gait cycle for each foot. The black dashed line represents the global wrist angle used as a threshold for detecting wrist vertical posture. The red solid dots represent the identified calibration reference points during the detected foot flat phases.  (c) The identified reference point for the right wrist based on the middle of the left foot flat phase during walking. (d) The identified reference point for the right wrist based on the middle of the right foot flat phase during walking.}
  \label{WalkingFFP}
  \end{center}
\end{figure*}

Where, $Threshold_{heel}$ and $Threshold_{fore}$ indicate the dynamic threshold used to determine if the heel area and the forefoot area are pressed, respectively. $n_{swing}$ indicates the number of samples in the swing phase when the foot is in the air and the flexible pressure sensor array is not pressed. In our previous research, a dynamic threshold has been proposed to identify the swing phase reliably \cite{chen2019bring}. The $HeelGRF$ and $ForeGRF$ indicate the ground reaction force (GRF) measured by the pressure sensors in the heel area and the forefoot area, respectively. $\sigma_{heel}$ and $\sigma_{fore}$ indicate the standard deviation of $HeelGRF$ and $ForeGRF$ during the swing phase. As shown in Fig. \ref{WalkingFFP} (A), the heel and forefoot regions of the flexible pressure sensor array are indicated with red and blue colors. Although the flexible pressure sensor array is not actively pressed during the swing phase, small contact forces exist and the measured $HeelGRF$ and $ForeGRF$ are nonzero. We hypothesize that the distribution of the $HeelGRF$ and $ForeGRF$ during the swing phase fit Gaussian distribution. Therefore, if the measured $HeelGRF$ and $ForeGRF$ are higher than the $Threshold_{heel}$ and $Threshold_{fore}$, respectively, the foot flat phase is recognized.



Similar to the process of recognizing the wrist vertical posture during standing, only when the wrist pitch angle is higher than the $Threshold_{angle}$, the wrist posture is recognized in a suitable ``vertical'' posture. However, during the foot flat phase, the wrist angle can be changed continuously in a range and there could be many candidate samples. To identify a suitable candidate sample as the KVLU reference point, the first and last sample above the $Threshold_{angle}$ was selected and only the one closest to the middle of the foot flat phase was used as the KVLU reference point for that gait cycle. This is because the supporting leg is more straight near the middle of the foot-flat phase (Fig. \ref{WalkingFFP}. (B)).

\subsubsection{Estimate the wrist vertical location at the KVLU reference point}
After the identification of KVLU reference points, accurately estimating their vertical location is crucial for calibrating the LVL measurement.
According to anthropometry, there exists a ratio between the length of different body segments and the body height. Therefore, the vertical location of the wrist at the KVLU reference points, where the body segments such as legs, trunk, and arms are straight, can be estimated with the known body height and a known ratio. In this research, data collected from a group of ten subjects was used to calculate the ratio between the wrist height at the KVLU reference point and body heights. The averaged ratio across ten subjects was used for estimating the wrist vertical location at the KVLU reference point by multiplying it with the known body height.



\subsubsection{LVL calibration with the KVLU reference point}
It is well known that the load vertical travel distance and the corresponding changes in the measured air pressure have a linear relationship \cite{ghoreishi2023barometer}. Therefore, with the KVLU reference point as a reference, the relationship between the load vertical travel distance and the air pressure changes can be described with Eq. \ref{eq_LVL1}.

\begin{equation} 
LVL - LVL_{KVLU} = a * (P - P_{KVLU}) + b
\label{eq_LVL1}
\end{equation}

Where, $LVL - LVL_{KVLU}$ indicates the load vertical travel distance relative to the KVLU reference point; $P - P_{KVLU}$ indicate the air pressure changes relative to the KVLU reference points; $a$ and $b$ indicate the parameters determined by the linear relationship between the load vertical travel distance and the corresponding changes in the measured air pressure. 

After transformation, Eq. \ref{eq_LVL2} can be used to estimate LVL in real-time. 

\begin{equation} 
LVL = a * (P - P_{KVLU}) + LVL_{KVLU} + b
\label{eq_LVL2}
\end{equation}

Since environmental-induced drifting errors have same impact on $P$ and $P_{KVLU}$, $LVL$ can be calibrated by updating the latest air pressure at the KVLU reference point ($P_{KVLU}$) whenever KVLU reference points are identified. 
To further calibrate LVL between the identification of KVLU reference points, the drifting errors between KVLU reference points will be estimated based on a popularly used linear drifting model \cite{li2023adaptive}. 

\section{Experiment and Results}

\subsection{Performance of the wearable system in identifying KVLU reference points and estimating the known vertical location during standing and walking}

\subsubsection{Experiment Design}
To evaluate the performance of the proposed wearable system in identifying KVLU reference points and estimating the known vertical location during standing and walking, an experiment was designed to involve 30 seconds of normal standing and three 60-second walking sections at slow (1 mile/hour), normal (2 mile/hour), and fast (3 mile/hour) walking speeds, respectively. 

Ten subjects with a height distribution of 171.3 cm ± 4.5 cm (mean ± std) participated in this experiment. The body height of each subject was measured before the experiment. During the experiment, the subject wore a pair of smart insoles and a pair of smart wristbands to perform standing and walking activities. To provide a ground-truth measurement of the vertical location of the wristbands, LED markers were placed on the wristbands and near the bottom of the experiment shoes. An optical motion capture system (PhaseSpace Impulse X2E) equipped with 8 cameras was used to track the 3D location of all the markers. The vertical location of the wristband from the ground can be achieved by calculating the difference in vertical location between the markers on the wrists and the makers on the shoes. This study was approved by the Institutional Review Board of the University of New Hampshire.

\subsubsection{Results of KVLU Reference Point Detection During Standing and Walking}

\begin{table*}[]
\caption{Percentage of the walking Steps With a Detected KVLU Reference Point}
\label{KVLUPercentage}
\begin{center}
\begin{threeparttable}
\begin{tabular}{clcccccccccccc}
\hline
\toprule
\multicolumn{2}{c}{\multirow{2}{*}{\textbf{\begin{tabular}[c]{@{}c@{}}Subject \\ ID\end{tabular}}}} & \multicolumn{4}{c}{\textbf{\begin{tabular}[c]{@{}c@{}}Slow \\ Walking\end{tabular}}} & \multicolumn{4}{c}{\textbf{\begin{tabular}[c]{@{}c@{}}Normal\\ Walking\end{tabular}}} & \multicolumn{4}{c}{\textbf{\begin{tabular}[c]{@{}c@{}}Fast \\ Walking\end{tabular}}} \\
\multicolumn{2}{c}{} & \textbf{RF-RW} & \textbf{RF-LW} & \textbf{LF-RW} & \textbf{LF-LW} & \textbf{RF-RW} & \textbf{RF-LW} & \textbf{LF-RW} & \textbf{LF-LW} & \textbf{RF-RW} & \textbf{RF-LW} & \textbf{LF-RW} & \textbf{LF-LW} \\ \hline
\multicolumn{2}{c}{\textbf{1}} & 3.57 & 100 & 100 & 0 & 100 & 100 & 100 & 45.45 & 100 & 100 & 100 & 79.45 \\
\multicolumn{2}{c}{\textbf{2}} & 11.11 & 97.23 & 91.67 & 8.33 & 55.77 & 100 & 100 & 59.62 & 100 & 96.67 & 100 & 98.31 \\
\multicolumn{2}{c}{\textbf{3}} & 85.29 & 100 & 100 & 100 & 54.55 & 100 & 100 & 97.72 & 94.29 & 57.14 & 100 & 100 \\
\multicolumn{2}{c}{\textbf{4}} & 100 & 40.48 & 100 & 95.24 & 100 & 73.33 & 100 & 100 & 100 & 82.46 & 100 & 100 \\
\multicolumn{2}{c}{\textbf{5}} & 100 & 100 & 100 & 100 & 100 & 100 & 100 & 100 & 100 & 100 & 100 & 100 \\
\multicolumn{2}{c}{\textbf{6}} & 100 & 100 & 100 & 100 & 100 & 100 & 100 & 100 & 100 & 100 & 100 & 100 \\
\multicolumn{2}{c}{\textbf{7}} & 40.63 & 100 & 100 & 100 & 95.62 & 100 & 100 & 100 & 100 & 100 & 100 & 100 \\
\multicolumn{2}{c}{\textbf{8}} & 47.37 & 100 & 100 & 60.53 & 100 & 98.75 & 100 & 97.92 & 100 & 98.04 & 100 & 100 \\
\multicolumn{2}{c}{\textbf{9}} & 100 & 100 & 100 & 100 & 100 & 100 & 100 & 100 & 100 & 91.94 & 100 & 98.39 \\
\multicolumn{2}{c}{\textbf{10}} & 97.44 & 100 & 100 & 94.87 & 100 & 100 & 100 & 100 & 100 & 100 & 100 & 100 \\ \hline
\multicolumn{2}{c}{\textbf{Mean}} & 71.63 & 93.11 & 99.17 & 78.24 & 97.11 & 90.95 & 100 & 90.33 & 97.38 & 89.69 & 100 & 97.03 \\ \hline
\end{tabular}
{\raggedright Note: ``RF'' stands for ``Right Foot'', ``LF'' stands for ``Left Foot'', ``RW'' stands for ``Right Wrist'', and ``LW'' stands for ``Left Wrist''. \par}
\end{threeparttable}
\end{center}
\end{table*}

\begin{figure}[h!]
\begin{center}
  \includegraphics[width=0.4\textwidth]{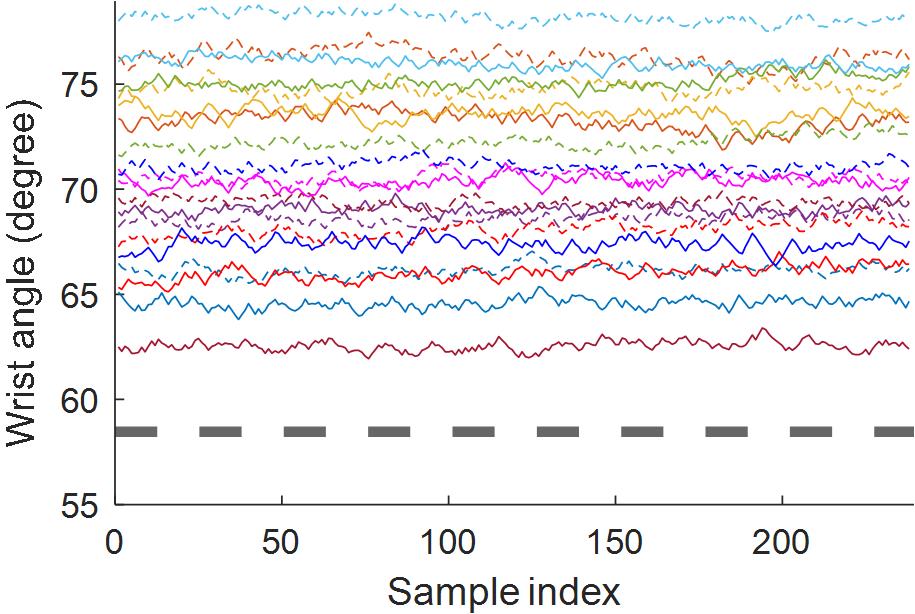}
  \caption{Wrist vertical posture identification with wrist angles during standing. The thick dashed black line indicates the $Threshold_{angle}$, and the plots with different colors indicate the wrist angle of different subjects during standing.}
  \label{KVLU_detection_standing}
  \end{center}
\end{figure}

During the standing experiment, all the subjects performed normal standing with legs straight and wrists in vertical position. Therefore, KVLU reference points exist in the standing of all the subjects. Fig. \ref{KVLU_detection_standing} shows the relationship between the wrist angles of different subjects during standing (indicated with thin curves of different colors) and the $Threshold_{angle}$ (indicated with the thick dashed black line). The $Threshold_{angle}$ calculated with the normal standing data of 10 subjects was 58.5 degrees. Since all wrist angles were higher than the $Threshold_{angle}$, the wrists of all the subjects were recognized as being in a vertical posture during the standing experiment. The standing activities were accurately recognized by the smart insole with the method proposed in \cite{chen2019bring}. Therefore, the KVLU reference points were accurately recognized for all the subjects during the standing experiment. 

Table \ref{KVLUPercentage} summarizes the percentage of walking steps with identified KVLU reference points in terms of different walking speeds and different combinations of feet and wrists. During walking, the KVLU reference points can be determined by the combination of different feet and wrists, including right foot and right wrist (RF-RW), right foot and left wrist (RF-LW), left foot and right wrist (LF-RW), and left foot and left wrist (LF-LW). To calibrate the LVL measurement of one wristband, two KVLU reference points can be used. For example, the KVLU reference points determined by RF-RW and LF-RW can be used to calibrate the LVL measured by the right wristband. Similarly, the KVLU reference points determined by RF-LW and LF-LW can be used to calibrate the LVL measured by the left wristband. For the result of each subject in Table \ref{KVLUPercentage}, when considering both reference points for each wrist, nearly 100\% of walking steps had qualified KVLU reference point(s) for each wristband. This result indicates that KVLU reference points can be frequently identified during walking for LVL calibration. 

In addition, the percentage of qualified KVLU reference points increases with walking speed. One possible reason for this observation is that the arm swings with a larger range and the wrist reaches a more vertical position when walking speed increases.

\begin{table*}[]
\caption{MAE of the estimated wrist vertical location at the detected KVLU reference points during walking}
\label{KVLUWalkingMAE}
\begin{center}
\begin{threeparttable}
\begin{tabular}{clcccccccccccc}
\hline
\toprule
\multicolumn{2}{c}{\multirow{2}{*}{\textbf{\begin{tabular}[c]{@{}c@{}}Subject \\ ID\end{tabular}}}} & \multicolumn{4}{c}{\textbf{\begin{tabular}[c]{@{}c@{}}Slow \\ Walking\end{tabular}}} & \multicolumn{4}{c}{\textbf{\begin{tabular}[c]{@{}c@{}}Normal\\ Walking\end{tabular}}} & \multicolumn{4}{c}{\textbf{\begin{tabular}[c]{@{}c@{}}Fast \\ Walking\end{tabular}}} \\
\multicolumn{2}{c}{} & \textbf{RF-RW} & \textbf{RF-LW} & \textbf{LF-RW} & \textbf{LF-LW} & \textbf{RF-RW} & \textbf{RF-LW} & \textbf{LF-RW} & \textbf{LF-LW} & \textbf{RF-RW} & \textbf{RF-LW} & \textbf{LF-RW} & \textbf{LF-LW} \\ \hline
\multicolumn{2}{c}{\textbf{1}} & 0.22 & 1.88 & 4.52 & NaN & 1.58 & 0.65 & 4.25 & 1.46 & 5.15 & 1.13 & 2.51 & 1.32 \\
\multicolumn{2}{c}{\textbf{2}} & 2.11 & 1.20 & 2.90 & 3.34 & 3.57 & 1.78 & 1.74 & 6.52 & 2.65 & 3.15 & 0.93 & 7.80 \\
\multicolumn{2}{c}{\textbf{3}} & 1.93 & 0.85 & 0.47 & 1.51 & 4.64 & 3.85 & 4.21 & 4.96 & 2.71 & 12.26 & 13.98 & 3.73 \\
\multicolumn{2}{c}{\textbf{4}} & 6.24 & 1.04 & 3.47 & 4.88 & 6.69 & 0.57 & 3.57 & 6.12 & 7.82 & 0.98 & 2.63 & 6.51 \\
\multicolumn{2}{c}{\textbf{5}} & 0.54 & 0.96 & 0.51 & 1.65 & 1.32 & 2.14 & 0.52 & 3.35 & 1.60 & 2.70 & 1.06 & 4.67 \\
\multicolumn{2}{c}{\textbf{6}} & 1.96 & 3.67 & 3.42 & 2.41 & 2.10 & 3.87 & 1.80 & 3.66 & 3.95 & 4.68 & 2.76 & 6.09 \\
\multicolumn{2}{c}{\textbf{7}} & 0.92 & 0.78 & 0.86 & 0.72 & 1.30 & 0.51 & 0.61 & 1.79 & 3.18 & 0.88 & 1.30 & 4.14 \\
\multicolumn{2}{c}{\textbf{8}} & 2.43 & 0.86 & 0.99 & 2.43 & 4.54 & 2.42 & 0.88 & 4.77 & 5.76 & 4.15 & 2.37 & 6.49 \\
\multicolumn{2}{c}{\textbf{9}} & 2.74 & 1.52 & 1.33 & 1.0 & 2.17 & 1.92 & 0.56 & 1.03 & 3.12 & 3.93 & 2.46 & 2.71 \\
\multicolumn{2}{c}{\textbf{10}} & 3.24 & 1.76 & 2.35 & 2.20 & 4.48 & 2.23 & 1.72 & 2.0 & 6.26 & 1.34 & 0.53 & 0.59 \\ \hline
\multicolumn{2}{c}{\textbf{Mean}} & 2.73 & 1.47 & 2.03 & 2.13 & 3.11 & 2.04 & 1.91 & 3.51 & 4.18 & 3.0 & 3.35 & 4.38 \\ \hline
\end{tabular}
{\raggedright Note: The unit of MAE is cm. \par}
\end{threeparttable}

\end{center}
\end{table*}

\subsubsection{Accuracy of the Estimated Wrist Vertical Location at the Detected KVLU Reference Points}


The height of the wrist vertical location at the detected KVLU reference points was estimated by multiplying the known body height with the ratio of wrist height and body height. With the measured ground truth height of the wrist vertical location during normal standing and the body height of ten subjects, the average ratio between the wrist height and body height was calculated to be 0.495. The accuracy of the estimated wrist vertical location at the detected KVLU reference points during walking was summarized in Table \ref{KVLUWalkingMAE}. 
The mean MAE of the estimated wrist vertical location at four different types of KVLU reference points determined by the combination of different feet and wrists ranged from 1.47 cm to 4.38 cm. These results indicate that the estimated wrist vertical location at the KVLU reference points detected during walking has a good accuracy for calibrating drifting errors which could cause an error of tens of centimeters or more in LVL measurement. 

One important finding from Table \ref{KVLUWalkingMAE} was that the estimated wrist height has a higher accuracy at the KVLU reference points determined by the wrist and foot on the opposite sides (i.e. RF-LW and LF-RW) across different walking speeds. Compared with the MAE achieved by the wrist and foot on the same side (i.e. RF-RW and LF-LW), the MAE achieved by the wrist and foot on the opposite sides was decreased by 37\%, 44\%, 23\% for slow, normal, and fast walking, respectively. In addition, the combination of opposite feet and wrists had a generally higher possibility of having qualified KVLU reference points, as shown in \ref{KVLUPercentage}. These results indicate that the KVLU reference points determined by the opposite feet and wrists have better performance in both accuracy and frequency of occurrence. 

The variability in accuracy at different walking speeds might be caused by the increase of wrist swing speed and range with the increase of walking speeds. In addition, the ``NAN'' in Table \ref{KVLUWalkingMAE} indicates that no qualified KVLU reference points were detected. 

The accuracy of the estimated wrist vertical location at the detected KVLU reference points during standing was summarized in Table \ref{KVLUStandingMAE}. 
The MAE of the estimated wrist height at the KVLU reference points during the standing activity was 2.04 cm and the corresponding accuracy was 97.47\%. This result indicates that the estimated wrist vertical location at the KVLU reference points detected during standing has good accuracy for calibrating drifting errors in LVL measurement.

\begin{table}[h!]
\caption{Accuracy of the Wrist vertical location estimation at the KVLU reference point during standing}
\label{KVLUStandingMAE}
\begin{center}
\begin{threeparttable}
\begin{tabular}{cc c c c c}
\hline
    \toprule
    \multicolumn{2}{c}{\textbf{\begin{tabular}[c]{@{}c@{}}Subject\\ID\end{tabular}}}  &
  \textbf{\begin{tabular}[c]{@{}c@{}} Estimated\\wrist\\ Height \\(cm)\end{tabular}} &
  \textbf{\begin{tabular}[c]{@{}c@{}} True\\ wrist \\ Height \\ (cm)\end{tabular}} &
  \textbf{\begin{tabular}[c]{@{}c@{}} MAE\\ (cm) \end{tabular}} &
  \textbf{\begin{tabular}[c]{@{}c@{}} Accuracy\\ (\%) \end{tabular}} \\
\hline
  
\multicolumn{2}{c}{\textbf{1}} & 82.56 & 80.23 & 2.34 & 97.08 
 \\ 
\multicolumn{2}{c}{\textbf{2}}    &  84.35 & 82.33 & 2.01 & 97.56 
 \\ 
\multicolumn{2}{c}{\textbf{3}}  & 84.35 & 83.70 & 0.65 & 99.22 
\\ 
\multicolumn{2}{c}{\textbf{4}} & 83.60 & 78.89 & 4.72 & 94.02 
 \\ 
\multicolumn{2}{c}{\textbf{5}}  & 83.60 & 82.04 & 1.45 & 98.23 
 \\ 
\multicolumn{2}{c}{\textbf{6}} & 83.85 & 79.61 & 4.24 & 94.67 
 \\ 
\multicolumn{2}{c}{\textbf{7}}  & 89.29 & 89.35 & 0.05 & 99.94 
 \\ 
\multicolumn{2}{c}{\textbf{8}} & 85.34 & 83.25 & 2.09 & 97.49 
 \\ 
\multicolumn{2}{c}{\textbf{9}} & 82.47 & 80.92 & 1.55 & 98.08
\\ 
\multicolumn{2}{c}{\textbf{10}} & 88.06 & 86.65 & 1.41 &98.37 
\\\hline 
\multicolumn{2}{c}{\textbf{Mean}} & 84.75 & 82.71 & 2.04 & 97.47 
\\
\hline 

\end{tabular}
\begin{tablenotes}

\end{tablenotes}
\end{threeparttable}
\end{center}

\end{table}

\subsection{Accuracy of the LVL measurement with KVLU calibration}


To evaluate the accuracy of the LVL measurement with KVLU calibration, an experiment was designed to involve lifting at various vertical locations. The KVLU calibration performance was evaluated with the standing activity only because the KVLU reference points can be identified accurately for both standing and walking. 

\begin{figure}[h]
\begin{center}
  \includegraphics[width=3in]{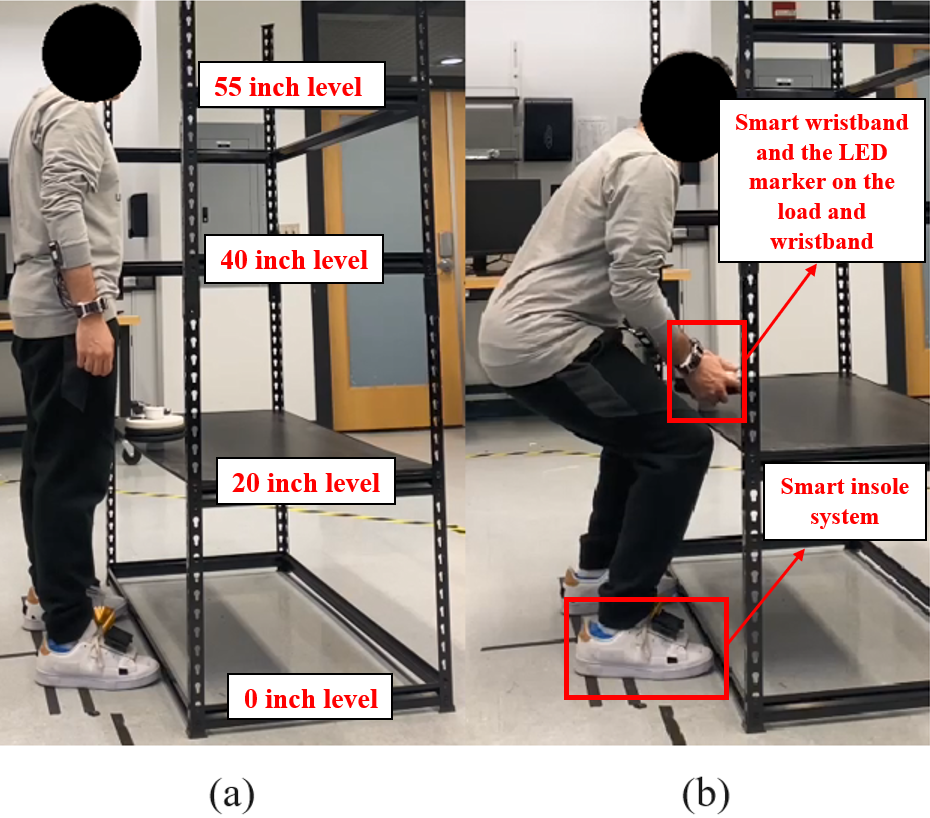}
  \caption{Experiment setup for evaluating the accuracy of the LVL measurement with KVLU calibration. (a) Normal standing posture with hands positioned vertically on the body sides; (b) Lifting posture with load in the hands.}
  \captionsetup{justification=centering}
  \label{StandingLifting}
  \end{center}
\end{figure}

As shown in Fig. \ref{StandingLifting}, vertical locations involved in the experiments included foot level (0 inches), knee level (20 inches), chest level (40 inches), and shoulder level (55 inches). 
During the experiment, the subject wore a pair of smart insoles and one smart wristband on the right wrist to perform 30 lifting activities at each vertical location. Only one smart wristband on the right wrist was used because the lifting activity involved in the experiment was popularly used double-hand symmetry lifting where the wrist vertical location of both wrists is similar. For each lifting activity, the subject lifted the load, held it for 3 seconds, put the load back, and returned to normal standing for 5 seconds. To provide a ground-truth measurement of the vertical location of the load, LED markers were placed on the load and tracked by an optical motion capture system (PhaseSpace Impulse X2E).

Table \ref{LiftingResults} summarizes the MAE between the measured wrist vertical location after calibration and the ground-truth LVL.
\begin{table}[h]

\caption{MAE Between the Estimated Wrist Vertical Location After Calibration and LVL during lifting (cm)}
\label{LiftingResults}
\begin{center}
\begin{threeparttable}
\begin{tabular}{cc c c c c c}
\hline
   \toprule
    \multicolumn{2}{c}{\textbf{\begin{tabular}[c]{@{}c@{}}Subject\\ID\end{tabular}}} &
  \textbf{\begin{tabular}[c]{@{}c@{}}Ground\\ Level\end{tabular}} &
  \textbf{\begin{tabular}[c]{@{}c@{}}Knee\\ Level\end{tabular}} &
  \textbf{\begin{tabular}[c]{@{}c@{}}Waist \\ Level\end{tabular}} &
  \textbf{\begin{tabular}[c]{@{}c@{}}Shoulder \\ Level\end{tabular}} &
  \textbf{\begin{tabular}[c]{@{}c@{}}Mean\end{tabular}} \\ \hline
\multicolumn{2}{c}{\textbf{1}}   & 5.55 & 5.70 & 10.09 & 4.03  & 6.10 
 \\ 
\multicolumn{2}{c}{\textbf{2}}    & 4.65 & 5.44 & 5.70 & 6.18 & 5.52 
 \\ 
 \multicolumn{2}{c}{\textbf{3}}    & 6.02 & 8.08 & 6.82 & 9.08  & 7.57 
 \\ 
\multicolumn{2}{c}{\textbf{4}}    & 4.04 & 3.21 & 10.34 & 6.68 & 5.96  
\\ 
\multicolumn{2}{c}{\textbf{5}}   & 5.15 & 4.39 & 4.75 & 3.84  & 4.53 
 \\ 
\multicolumn{2}{c}{\textbf{6}}   & 3.22 & 3.72 & 5.28 & 7.09 & 4.84 
 \\ 
\multicolumn{2}{c}{\textbf{7}}    & 2.82 & 3.86 & 3.99 & 3.91  & 3.69  
 \\ 
\multicolumn{2}{c}{\textbf{8}}    & 2.51 & 3.14 & 5.19 & 10.51 & 5.57 
 \\ 
\multicolumn{2}{c}{\textbf{9}}    & 4.94 & 4.98 & 4.49 & 5.38 & 4.95 
\\ 
\multicolumn{2}{c}{\textbf{10}}  & 4.05 & 5.35 & 10.31 & 14.17  & 8.68 \\ 

 \hline
\multicolumn{2}{c}{\textbf{Mean}}  & 4.31 & 4.76 & 6.6 & 7.22 & 5.71 \\\hline 
 
\end{tabular}
\begin{tablenotes}

\end{tablenotes}
\end{threeparttable}
\end{center}

\end{table}

The MAE at ground-level, knee-level, waist-level, and shoulder-level were 4.31, 4.76, 6.60, and 7.22 cm, respectively. The overall MAE of the wearable sensor system for measuring LVL was 5.71 cm. According to the manual of the RNLE, a 3.3 cm change in LVL leads to 1\% changes in the lifting index, which is an indicator of the lifting risk \cite{waters1994applications}. Therefore, a measurement error of 5.71 cm is expected to lead to a 1.73\% (5.71/3.3*1\%) error in the lifting risk assessment. 

\begin{figure}[h]
\begin{center}
    \includegraphics[width=3in]{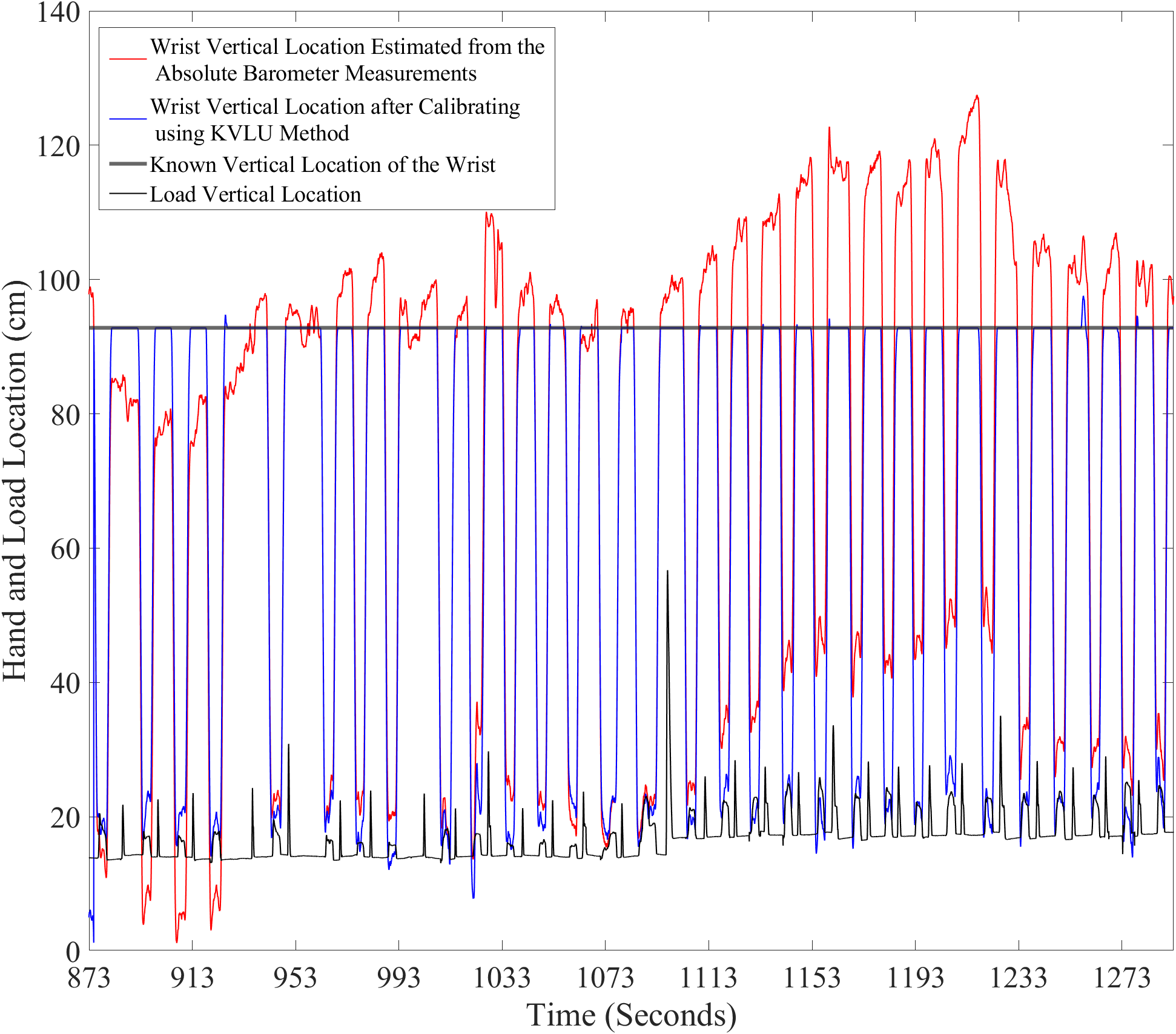}
    \caption{LVL measurement during ground-level lifting before and after applying the KVLU calibration}
    \label{fig:LVLAfterCalibration}
\end{center}
\end{figure}

To visualize the performance of the KVLU in correcting the environmental-induced LVL measurement errors, Fig. \ref{fig:LVLAfterCalibration} shows the wrist vertical location calculated with the raw barometric pressure (red line), the wrist vertical location calibrated with KVLU (blue line), and the ground-truth LVL (black line) during the lifting at the ground level. Please note, the LVL was not zero because the load was placed on a plate with a height to facilitate lifting.
As shown in Fig. \ref{fig:LVLAfterCalibration}, the proposed KVLU method significantly improve the accuracy and reliability of the LVL measurement.  

\vspace{-2mm}

\section{Discussion}
\subsection{Improvement of the Proposed Wearable LVL Measurement System Over State-of-the-Art Solutions} Experiment results showed that the proposed wearable system and the KVLU calibration method can measure LVL with good accuracy for assessing the lifting hazards reliably. Compared with existing studies, the wearable system proposed in this study showed a significant improvement in the accuracy of LVL measurement. As shown in Table \ref{CurrentMethodsComparison}, the LVL measurement error of our system (5.71 cm) is much smaller than that reported in existing studies (33cm and 25 cm for \cite{barim} and \cite{hlucny2020characterizing}, respectively). According to the RNLE, the 33 cm and 25 cm errors in LVL measurement will lead to 10\%  and 7.6\% errors in the lifting risk assessment \cite{waters1994applications}. The method proposed in this study can significantly reduce the errors in lifting risk assessment to 1.73\%. Additionally, to the best of our knowledge, the proposed wearable system in this study is the only system that has the capability for auto-calibration which is critical for ensuring accurate and long-term measurement in uncontrolled workplace environments. 

Compared with existing research, the unobtrusive wearable system proposed in this study only consists of a pair of smart insoles and up to a pair of smart wristbands. Since insoles are everyday used items and wristbands are well accepted, this wearable system design is expected to have enhanced user acceptance than wearable systems that need to attach numerous sensing devices to body segments like the thigh, where people usually don't attach additional items. 

\begin{table}[h!]
    \caption{Comparison with existing LVL measurement Methods}
    \label{CurrentMethodsComparison} 
    \begin{center}
\begin{threeparttable}
\begin{tabular}[]{cccccccc}
\hline
\toprule
       \multicolumn{2}{c}{\multirow{3}{*}{\textbf{\begin{tabular}[c]{@{}c@{}} Study \end{tabular}}}}  &
       \multicolumn{2}{c}{\multirow{3}{*}{\textbf{\begin{tabular}[c]{@{}c@{}} Wearable \\ System \end{tabular}}}} &
       \multicolumn{2}{c}{\multirow{3}{*}{\textbf{\begin{tabular}[c]{@{}c@{}} Auto \\ Calibration \end{tabular}}}}  &
       \multicolumn{2}{c}{\multirow{3}{*}{\textbf{\begin{tabular}[c]{@{}c@{}} Measurement \\ Error \end{tabular}}}} \\
       \\
       \\
       \hline
       \multicolumn{2}{c}{\multirow{3}{*}{\textbf{\begin{tabular}[c]{@{}c@{}} \cite{hlucny2020characterizing} \end{tabular}}}}  
       & \multicolumn{2}{c}{\multirow{3}{*}{{\begin{tabular}[c]{@{}c@{}} 17 IMU sensors \\attached to different\\ body segments \end{tabular}}}}
       & \multicolumn{2}{c}{\multirow{3}{*}{{\begin{tabular}[c]{@{}c@{}} $\times$ \end{tabular}}}}
       & \multicolumn{2}{c}{\multirow{3}{*}{{\begin{tabular}[c]{@{}c@{}} 25 cm \end{tabular}}}}\\
       \\
       \\ \\
       \multicolumn{2}{c}{\multirow{3}{*}{{\begin{tabular}[c]{@{}c@{}} \cite{barim} \end{tabular}}}}  
       & \multicolumn{2}{c}{\multirow{3}{*}{{\begin{tabular}[c]{@{}c@{}} 5 IMU  sensors \\ attached to  thighs, \\arms, and waist \end{tabular}}}}
       & \multicolumn{2}{c}{\multirow{3}{*}{\textbf{\begin{tabular}[c]{@{}c@{}} $\times$ \end{tabular}}}}
       & \multicolumn{2}{c}{\multirow{3}{*}{{\begin{tabular}[c]{@{}c@{}} 33 cm \end{tabular}}}}\\
           \\
       \\
       \\
       
       \multicolumn{2}{c}{\multirow{3}{*}{{\begin{tabular}[c]{@{}c@{}} Our \end{tabular}}}}  
       & \multicolumn{2}{c}{\multirow{3}{*}{{\begin{tabular}[c]{@{}c@{}} 2 smart insoles and \\2 smart wristbands \end{tabular}}}}
       & \multicolumn{2}{c}{\multirow{3}{*}{{\begin{tabular}[c]{@{}c@{}} KVLU \end{tabular}}}}
       & \multicolumn{2}{c}{\multirow{3}{*}{{\begin{tabular}[c]{@{}c@{}} 5.71 cm \end{tabular}}}}\\  
       \\
       \\
       \hline
    
\end{tabular}
\end{threeparttable}
\end{center}
\end{table}

\subsection{The Error Introduced by Using Wrist Vertical Location as the LVL}
In this section, the MAE and mean error (ME) of using wrist vertical location as the LVL were summarized in Table \ref{LiftingResults3} and  Table \ref{LiftingResultswithSTD3}, respectively, to show the error distribution across different load height levels. As shown in Table \ref{LiftingResultswithSTD3}, the ME at the waist-level load height is near zero (-0.39 cm), but shifts in the negative direction to -1.77 at the shoulder-level. Conversely, it changes in a positive direction to 4.1 cm and 5.91 cm at the knee-level and ground-level, respectively. These trends occur because when lifting loads at the waist level, the wrist is held horizontally and the vertical location of the wrist and the load are similar; when lifting loads from lower levels, the wrist position tends to be above the load (positive ME); when lifting loads at higher levels, the wrist position tends to be below the load (negative ME). Although the wrist vertical location is not an exact representation of the LVL, it can be used to track the LVL with a good accuracy. As shown in Table \ref{LiftingResults3}, the average MAE is only 3.69 cm. These results showed that it is feasible to measure LVL with the wrist vertical location. In the future, we will further improve the accuracy of the LVL measurement by using the observed error distribution across various load height levels.


\begin{table}[h]
\caption{MAE Between Ground-Truth Wrist Vertical Location and LVL during Lifting (cm)}
\label{LiftingResults3}
\begin{center}
\begin{threeparttable}
\begin{tabular}{cc c c c c c}
\hline
   \toprule
    \multicolumn{2}{c}{\textbf{\begin{tabular}[c]{@{}c@{}}Subject\\ID\end{tabular}}} &
  \textbf{\begin{tabular}[c]{@{}c@{}}Ground\\ Level\end{tabular}} &
  \textbf{\begin{tabular}[c]{@{}c@{}}Knee\\ Level\end{tabular}} &
  \textbf{\begin{tabular}[c]{@{}c@{}}Waist \\ Level\end{tabular}} &
  \textbf{\begin{tabular}[c]{@{}c@{}}Shoulder \\ Level\end{tabular}} &
  \textbf{\begin{tabular}[c]{@{}c@{}}Mean\end{tabular}} \\ \hline
\multicolumn{2}{c}{\textbf{1}}   & 5.38  & 3.79  & 1.69  & 2.50   & 3.38 
 \\ 
\multicolumn{2}{c}{\textbf{2}}    & 8.76  & 6.39  & 2.28  & 2.79  & 5.09 
 \\ 
 \multicolumn{2}{c}{\textbf{3}}    & 6.95  & 3.33  & 1.94  & 2.53  & 3.87 
 \\ 
\multicolumn{2}{c}{\textbf{4}}    & 5.37 & 4.39 & 1.25  & 2.14  & 3.32  
\\ 
\multicolumn{2}{c}{\textbf{5}}   & 5.65 & 3.46  & 1.81 & 2.13   & 3.28 
 \\ 
\multicolumn{2}{c}{\textbf{6}}   & 6.01  & 3.26  & 2.34  & 2.88  & 3.65 
 \\ 
\multicolumn{2}{c}{\textbf{7}}    & 5.44  & 5.87  & 1.91  & 2.01  & 4.13  
 \\ 
\multicolumn{2}{c}{\textbf{8}}    & 7.82  & 6.13 & 2.89  & 2.71 & 4.91 
 \\ 
\multicolumn{2}{c}{\textbf{9}}    & 4.14  & 3.56  & 1.07  & 2.97  & 2.97 
\\ 
\multicolumn{2}{c}{\textbf{10}}  & 3.61  & 2.06  & 1.52 & 2.17  & 2.34 \\ 

 \hline
\multicolumn{2}{c}{\textbf{Mean}}  & 5.93 & 4.33 & 1.84 & 2.48 & 3.69 \\\hline 
 
\end{tabular}
\begin{tablenotes}

\end{tablenotes}
\end{threeparttable}
\end{center}
\end{table}

\begin{table}[h]
\caption{ME Between Ground-Truth Wrist Vertical Location and LVL during Lifting (cm)}
\label{LiftingResultswithSTD3}
\begin{center}
\begin{threeparttable}
\begin{tabular}{cc c c c c c}
\hline
    \toprule
    \multicolumn{2}{c}{\textbf{\begin{tabular}[c]{@{}c@{}}Subject\\ID\end{tabular}}} &
  \textbf{\begin{tabular}[c]{@{}c@{}}Ground\\ Level\end{tabular}} &
  \textbf{\begin{tabular}[c]{@{}c@{}}Knee\\ Level\end{tabular}} &
  \textbf{\begin{tabular}[c]{@{}c@{}}Waist \\ Level\end{tabular}} &
  \textbf{\begin{tabular}[c]{@{}c@{}}Shoulder \\ Level\end{tabular}} &
  \textbf{\begin{tabular}[c]{@{}c@{}}Mean\end{tabular}} \\ \hline
\multicolumn{2}{c}{\textbf{1}}   & 5.35 (3.51) & 2.98 (3.61) & -1.14 (1.64) & -1.70 (2.53)  & 1.36 
 \\ 
\multicolumn{2}{c}{\textbf{2}}    & 8.76 (3.07) & 6.39 (2.43) & -1.53 (2.27) & -2.26 (2.75) & 2.82 
 \\ 
 \multicolumn{2}{c}{\textbf{3}}    & 6.95 (3.06) & 2.45 (3.07) & 0.21 (2.10) & -1.48 (2.69)  & 2.22 
 \\ 
\multicolumn{2}{c}{\textbf{4}}    & 5.37 (1.91) & 4.39 (1.86) & -0.88 (1.36) & -1.11 (2.32) & 1.98  
\\ 
\multicolumn{2}{c}{\textbf{5}}   & 5.65 (2.23) & 3.46 (1.86) & -1.33 (1.84) & -0.87 (2.66)  & 1.78 
 \\ 
\multicolumn{2}{c}{\textbf{6}}   & 6.01 (3.14) & 3.09 (2.22) & -0.87 (2.33) & -2.55 (2.61) & 1.45 
 \\ 
\multicolumn{2}{c}{\textbf{7}}    & 5.38 (2.99) & 5.86 (2.87) & 0.39 (2.19) & -1.39 (2.27)  & 3.0  
 \\ 
\multicolumn{2}{c}{\textbf{8}}    & 7.82 (2.02) & 6.13 (1.54) & 2.54 (2.34) & -1.72 (2.89) & 3.53 
 \\ 
\multicolumn{2}{c}{\textbf{9}}    & 4.05 (2.55) & 3.26 (2.82) & -0.41 (1.34) & -2.91 (2.45) & 1.06 
\\ 
\multicolumn{2}{c}{\textbf{10}}  & 3.61 (2.36) & 1.75 (1.86) & -0.65 (1.74) & -1.76 (2.23)  & 0.65 \\ 

 \hline
\multicolumn{2}{c}{\textbf{Mean}}  & 5.91 & 4.10 & -0.39 & -1.77 & 2.01 \\\hline 
 
\end{tabular}
\begin{tablenotes}

\end{tablenotes}
{\raggedright Note: in the parenthesis is the standard deviation. \par}
\end{threeparttable}
\end{center}

\end{table}

\section{Conclusion}

In this research, an unobtrusive wearable system was developed for measuring LVL reliably in uncontrolled workplace environments. Different from widely used IMU-based solutions, barometer-based LVL measurement was implemented and its advantages were demonstrated. A novel KVLU calibration method was proposed to enable reliable LVL measurement over long periods. The experiment results showed that the proposed LVL measurement system achieved an overall MAE of 5.71 cm. This result indicates that the proposed system can ensure reliable LVL measurement, enhance lifting risk assessment, and contribute to LBP prevention in uncontrolled workplace environments.

\bibliographystyle{IEEEtran} 
\bibliography{refs} 
\end{document}